\documentclass[graybox]{SNmult}

\usepackage{type1cm}		

\usepackage{makeidx}		 
\usepackage{graphicx}		
\usepackage{multicol}		
\usepackage[bottom]{footmisc}
\usepackage{newtxtext}	   
\usepackage[varvw,varbb]{newtxmath}	   

\usepackage[utf8]{inputenc}
\usepackage[T1]{fontenc} 
\usepackage[english]{babel}
\usepackage{fix-cm} 
\usepackage{amsthm,amsfonts} 
\usepackage{isotope}
\usepackage{braket}
\usepackage{siunitx}
\DeclareSIUnit\torr{Torr} 
\DeclareSIUnit\gauss{G}
\DeclareSIUnit\atoms{atoms}

\makeatletter
\providecommand*{\toclevel@titlech}{0} 
\edef\toclevel@authorch{\the\numexpr\toclevel@titlech+1}
\providecommand*{\toclevel@titlech}{0}
\def\toclevel@authorch{1000}
\makeatother

\usepackage{hyperref}
\usepackage[hyperpageref]{backref} 
\hypersetup{
    colorlinks=true,
    linkcolor=blue,
    filecolor=blue,      
    urlcolor=cyan,
    citecolor=blue,
    pdftitle={The Legacy of Enrico Fermi to Varenna},
    pdfauthor={Vladislav Gavryusev and Massimo Inguscio},
    pdfpagemode=FullScreen,
    }
    
\makeindex             

\begin{document}

%
%
%

\title*{The Legacy of Enrico Fermi to Varenna}
\author{Vladislav Gavryusev\orcidID{0000-0001-7734-7828} and\\ 
Massimo Inguscio\orcidID{0000-0001-8152-8103}}
\tocauthor{Vladislav Gavryusev and Massimo Inguscio}
\institute{Vladislav Gavryusev \at Department of Physics and Astronomy, University of Florence, Via G. Sansone 1, 50019, Sesto Fiorentino, Italy, 
and \at European Laboratory for Non-Linear Spectroscopy (LENS), University of Florence, Via N. Carrara 1, 50019, Sesto Fiorentino, Italy,
\email{vladislav.gavryusev@unifi.it}
\and 
Massimo Inguscio \at European Laboratory for Non-Linear Spectroscopy (LENS), University of Florence, Via N. Carrara 1, 50019, Sesto Fiorentino, Italy,
and \at National Institute of Optics (CNR-INO), National Research Council, Via N. Carrara 1, 50019, Sesto Fiorentino, Italy, 
\email{inguscio@lens.unifi.it}}

\maketitle

\abstract*{The Varenna school is a hub where generations of physicists, including numerous Nobel laureates, have shaped the field, often through collaborative exchanges across political and cultural boundaries. We examine the scientific legacy of Enrico Fermi and its influence on modern atomic, molecular, and optical physics. Beginning with Fermi’s 1954 lectures at the Varenna school, key developments are traced from high-energy physics to laser spectroscopy, precision metrology, and the control of ultracold atoms. Milestones such as Doppler-free spectroscopy, optical frequency combs, Bose–Einstein condensation, and degenerate Fermi gases are highlighted as turning points leading to quantum simulation and quantum computation. Fermi’s early advocacy for building a computer, rather than buying it, can be viewed as a precursor to today’s efforts in quantum science and technologies. This historical trajectory and legacy continues to inform current research in quantum matter and information science.}

\abstract{The Varenna school is a hub where generations of physicists, including numerous Nobel laureates, have shaped the field, often through collaborative exchanges across political and cultural boundaries. We examine the scientific legacy of Enrico Fermi and its influence on modern atomic, molecular, and optical physics. Beginning with Fermi’s 1954 lectures at the Varenna school, key developments are traced from high-energy physics to laser spectroscopy, precision metrology, and the control of ultracold atoms. Milestones such as Doppler-free spectroscopy, optical frequency combs, Bose–Einstein condensation, and degenerate Fermi gases are highlighted as turning points leading to quantum simulation and quantum computation. Fermi’s early advocacy for building a computer, rather than buying it, can be viewed as a precursor to today’s efforts in quantum science and technologies. This historical trajectory and legacy continues to inform current research in quantum matter and information science.}

\section{Enrico Fermi and the development milestones of atomic physics marked through Varenna schools}
\label{Inguscio:sec:EF_Varenna}

It is a long tradition of the Varenna Schools to foster discussions between young and mature people which are fundamental for the progress of science. The inspiring avenue of Villa Monastero has hosted nearly one hundred future Nobel laureates, many of whom first attended as students before returning as distinguished lecturers. This school is focused on the field of quantum simulation and quantum computing with ultracold atoms, a fascinating topic which comes from long lasting developments in atomic and molecular physics.

\begin{figure}[htb]
	\centering
	\includegraphics[width=0.52\textwidth]{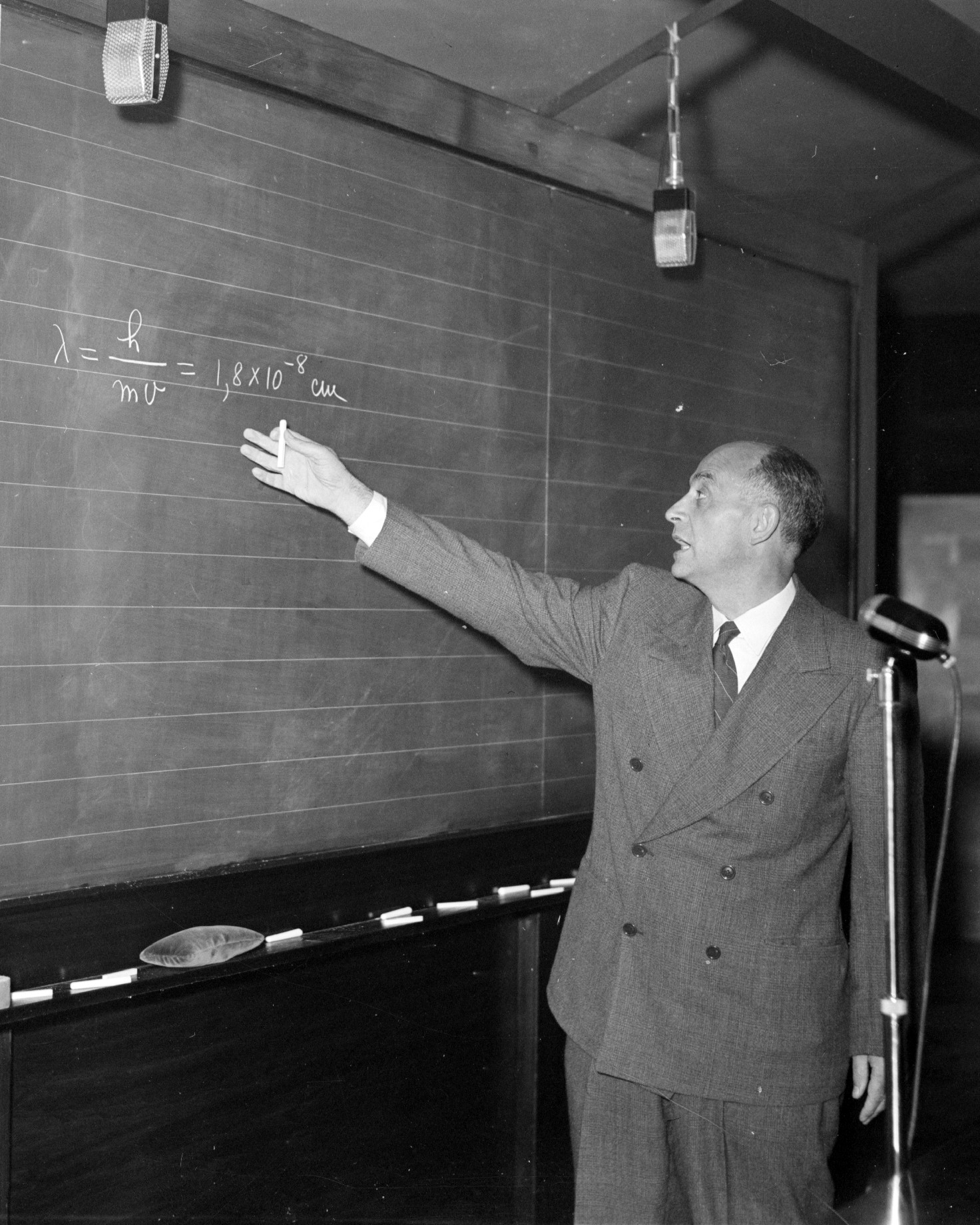}
	\caption{Enrico Fermi writing about the de Broglie length on the blackboard, on the 20th of October 1949, at a press conference at Montecatini in Milan. Photograph by Publifoto, copyright Intesa Sanpaolo Publifoto Archive, reproduced with permission.}
	\label{Inguscio:fig:Fermi}
\end{figure}

The title of this contribution, ``The Legacy of Enrico Fermi to Varenna'', comes both from the fact that this great scientist was an early lecturer in Varenna and that most of his ideas are still underpinning current developments in the scientific world. We would like to show how Fermi's legacy is related to the fields in which we all are involved, although he is better known as a nuclear physicist for which he received the Nobel Prize in 1938. He was here 70 years ago in July 1954, giving his first and last lecture at this school, which had been founded one year earlier and a few months before his death. As displayed in Figure~\ref{Inguscio:fig:Fermi}, he was teaching at this blackboard about the de Broglie length in the context of high energy physics, but this concept applies across all energy scales, including the atomic one, highlighting that there are no boundaries, nor due to energy or other factors.

The next year school of 1955 was directed by Isaac Rabi and Charles Townes was among the teachers. That was the beginning of high precision measurements of molecular spectroscopy and Rabi dedicated this school to Enrico Fermi who unfortunately had passed away at the end of 1954. Isaac was teaching about quadrupole transitions, and Charles was starting to present the inversion in ammonia. These topics are now basic, but still worthy of being studied, and they are part of the intriguing and entangled story of light and matter, lasers and atoms or molecules.

\begin{figure}[htb]
	\centering
	\includegraphics[width=\textwidth]{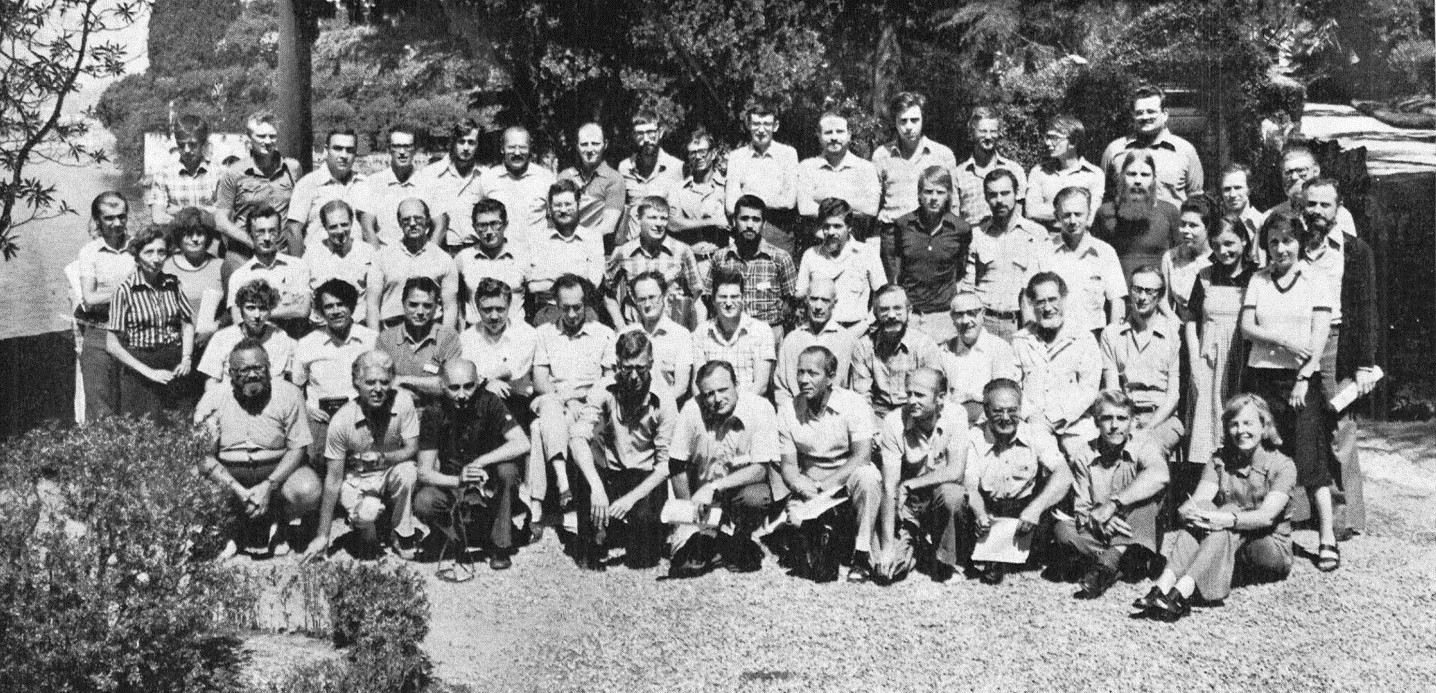}
	\caption{Group photo of the International School of Physics ``Enrico Fermi'' Course LXVIII on ``Metrology and Fundamental Constants'' that took place in Villa Monastero in Varenna on Lake Como between 12th and 24th July 1976. The course was directed by A. Ferro Milone and P. Giacomo, with the support of the scientific secretary S. Leschiutta. Copyright Societ\`a Italiana di Fisica (SIF), reproduced with permission.}
	\label{Inguscio:fig:course_LXVIII}
\end{figure}

The course LXVIII on ``Metrology and Fundamental Constants'' held in 1976, whose participants are shown in Figure~\ref{Inguscio:fig:course_LXVIII}, represented an important development in the history of precision metrology. This was the first Varenna school to which I, Massimo Inguscio, participated (as a young researcher then) and where I attended the lectures of Veniamin Chebotayev, a foundational figure in laser spectroscopy who specialized in using lasers to mitigate Doppler broadening, thereby allowing for the investigation of matter with unprecedented resolution~\cite{Chebotayev1980}. A defining characteristic of this period was the ``Science for Peace'' ethos. Despite the geopolitical constraints of the Cold War, the school successfully integrated researchers from across the Iron Curtain, including the Soviet Union, United States of America, Germany, and France. This international synthesis and shared approach underscored the role of scientific investigation as a unifying global force that transcended contemporary political barriers, supported peace and facilitated dialogue. 

\begin{figure}[htb]
	\centering
	\includegraphics[width=0.8\textwidth]{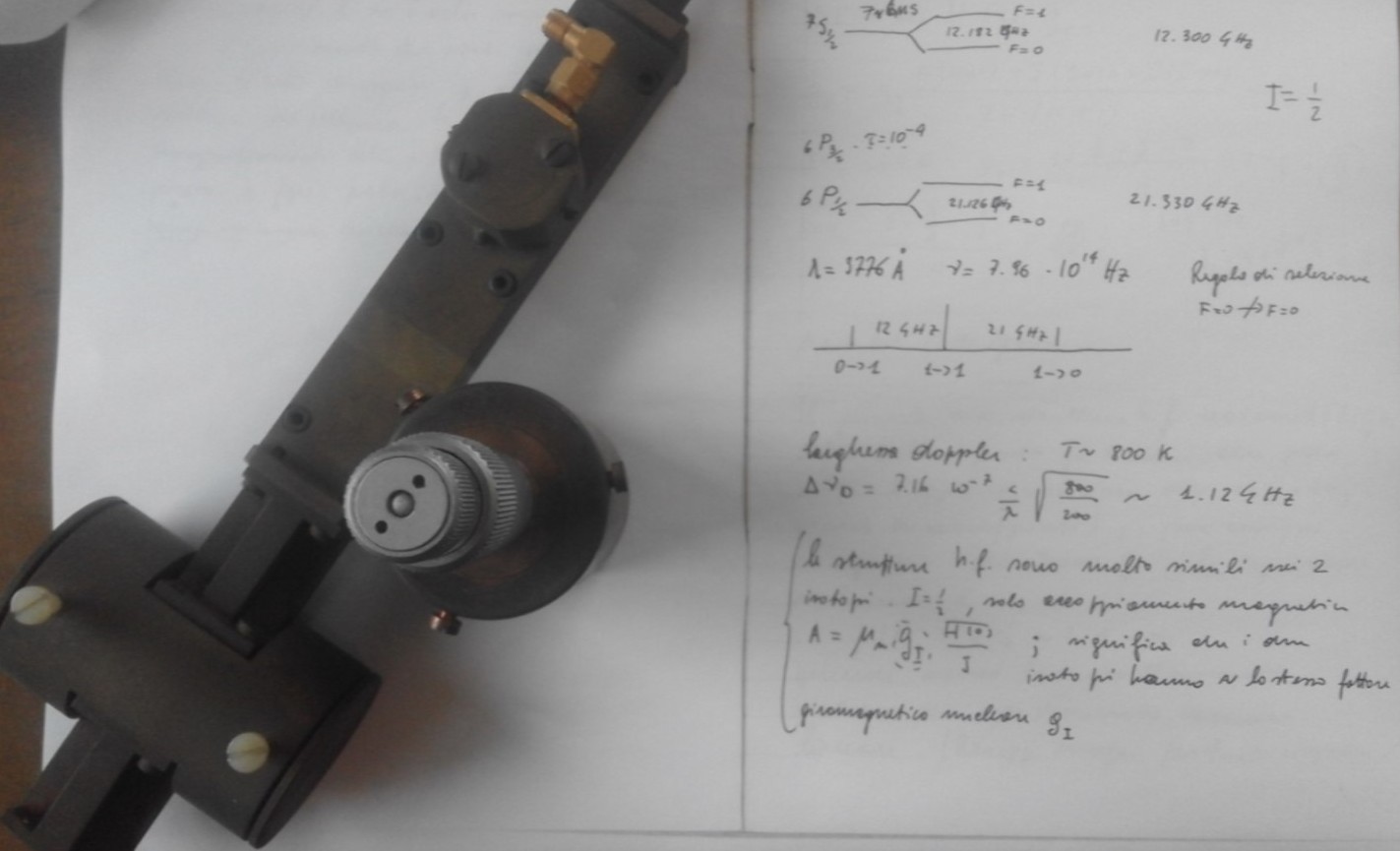}
	\caption{Handwritten notes by M. Inguscio on increasing the measurement precision for frequency metrology by moving from Caesium ($\qty{9.2}{\giga\hertz}$) to Thallium ($\qty{21.3}{\giga\hertz}$) in 1973, Istituto Elettrotecnico Nazionale Galileo Ferraris (now INRiM).}
	\label{Inguscio:fig:spettroscopia}
\end{figure}

The evolution of frequency metrology progressed from microwave-based Cesium standards toward higher-frequency optical regimes. Figure~\ref{Inguscio:fig:spettroscopia} captures the early experimental attempts by M. Inguscio to bridge this gap in 1973, specifically the transition toward Thallium standards. This trajectory from the microwave domain toward optical frequencies served as a necessary precursor to the development of optical clocks. 

\begin{figure}[htb]
	\centering
	\includegraphics[width=\textwidth]{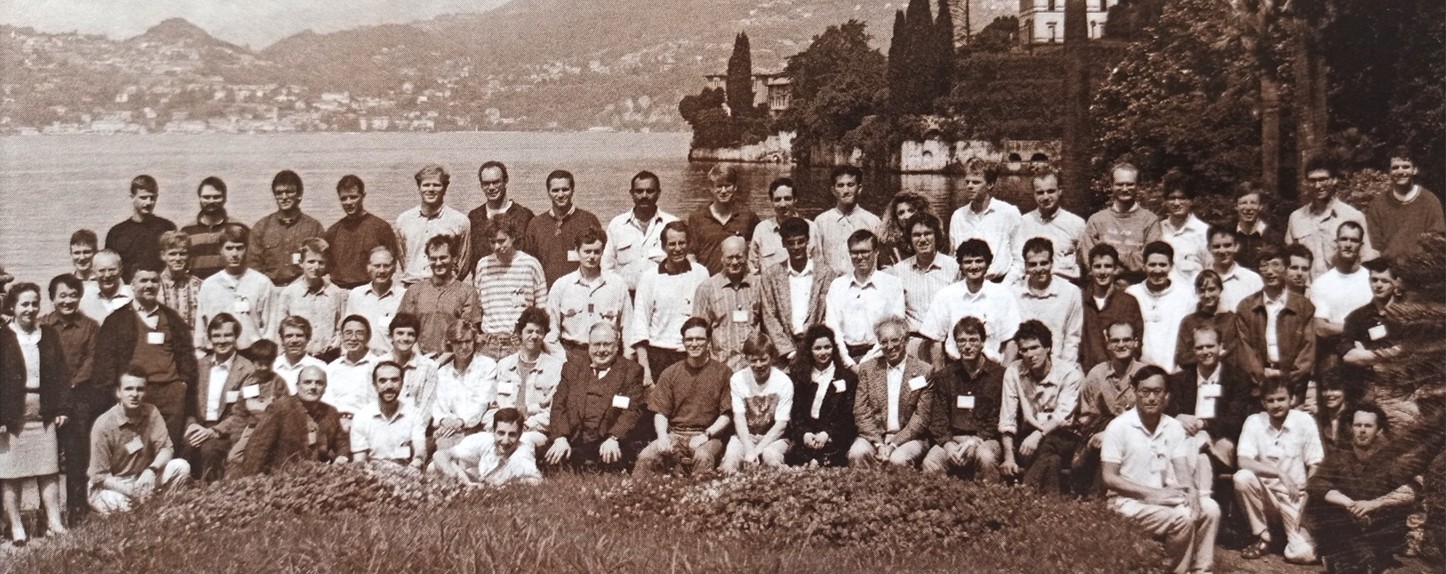}
	\caption{Group photo of the International School of Physics ``Enrico Fermi'' Course CXX on ``Frontiers in Laser Spectroscopy'' that took place in Villa Monastero in Varenna on Lake Como between 23rd June and 3rd July 1992. The course was directed by T.W. H\"ansch and M. Inguscio. Copyright Societ\`a Italiana di Fisica (SIF), reproduced with permission.}
	\label{Inguscio:fig:course_CXX}
\end{figure}

\begin{figure}[htb]
	\centering
	\includegraphics[width=0.7\textwidth]{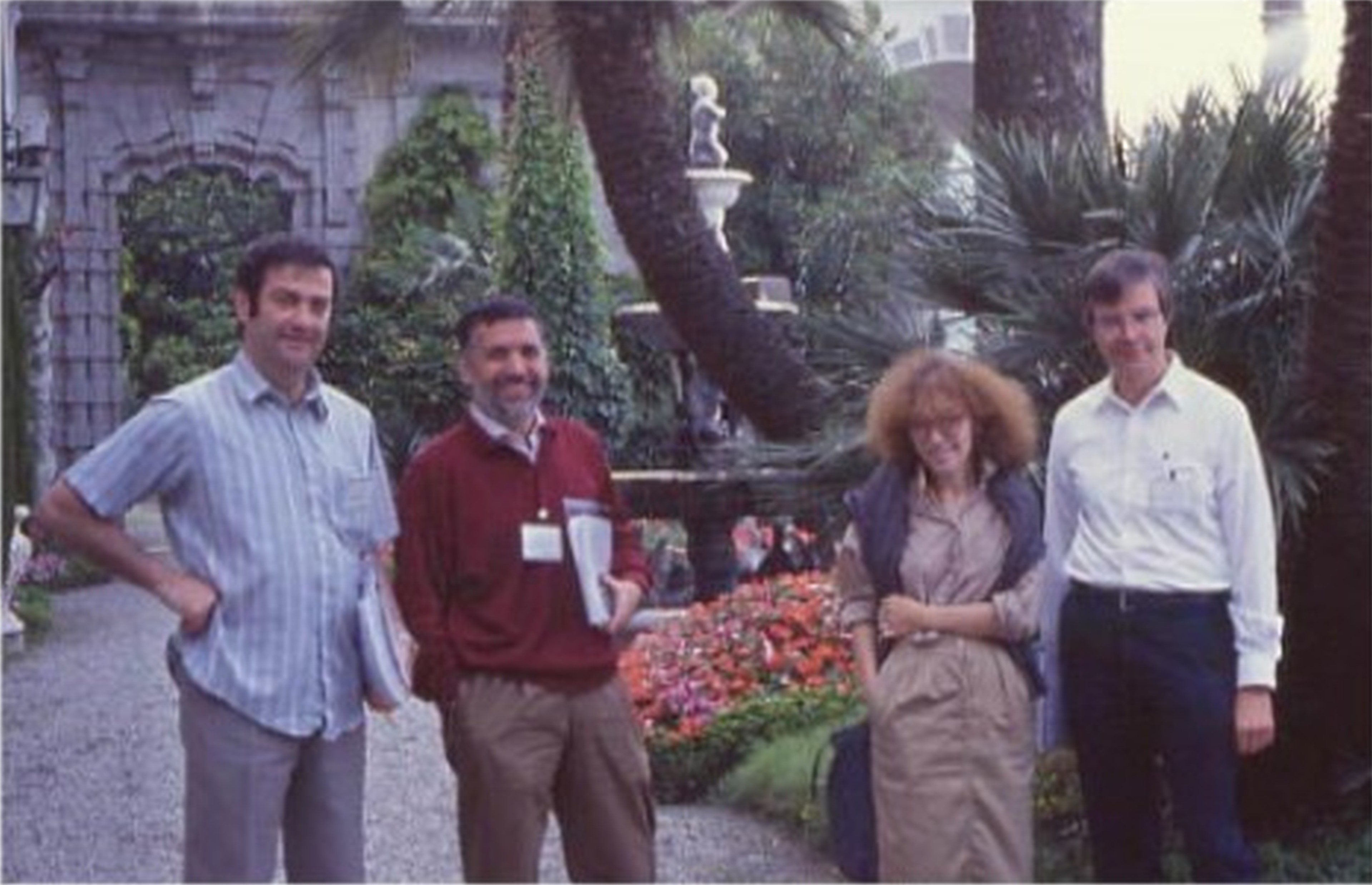}
	\caption{Photo of (from left to right) S. Haroche, M. Inguscio, C. Haroche and T. W. H\"ansch at the International School of Physics ``Enrico Fermi'' Course CXX on ``Frontiers in Laser Spectroscopy'' (1992).}
	\label{Inguscio:fig:course_CXX_HKN}
\end{figure}

The 1992 course CXX on ``Frontiers in Laser Spectroscopy'' (Figure~\ref{Inguscio:fig:course_CXX}) addressed the contemporary frontiers of matter spectroscopy and was attended by many young researchers, such as Steve Chu and Marco Bellini, and renowned experts like Ted W. H\"ansch, Arthur L. Schawlow and V. Chebotayev. In particular, it enabled frequent dialogues between H\"ansch and Chebotayev (who sadly died shortly after) and an ensuing collaboration between H\"ansch and Bellini~\cite{Bellini2000}. These two experiences stimulated Ted towards the eventual realization of the optical frequency comb~\cite{Udem1999}, a revolutionary and Nobel Prize-winning advancement in 2005~\cite{Haensch2006} for this field and many other areas. Another lecturer who greatly contributed to precision laser metrology is Kenneth M. Evenson through his efforts to determine the speed of light~\cite{Evenson1972} that is a foundational constant for modern metrology, now based on quantum mechanics. Today these scientific and technological efforts have allowed replacing cumbersome frequency chains with more efficient and much more compact laser systems. Moreover, these stories highlight that the open and welcoming social environment of Villa Monastero serves as a strong catalyst for inter-generational mentorship. The transfer of ideas through informal discussion emphasizes the critical role of human relationships in the scientific process, as they foster new intuitions, fruitful collaborations and friendships, including family members. Such outcomes are exemplified in Figure~\ref{Inguscio:fig:course_CXX_HKN} by the meeting between T. W. H\"ansch, M. Inguscio and Serge Haroche with his wife Claudine Haroche. Interestingly, Serge's studies of Rydberg atoms for high sensitivity detection and quantum non-demolition investigations later led him to receive the Nobel Prize in 2012~\cite{Haroche2013}.

\begin{figure}[htb]
	\centering
	\includegraphics[width=\textwidth]{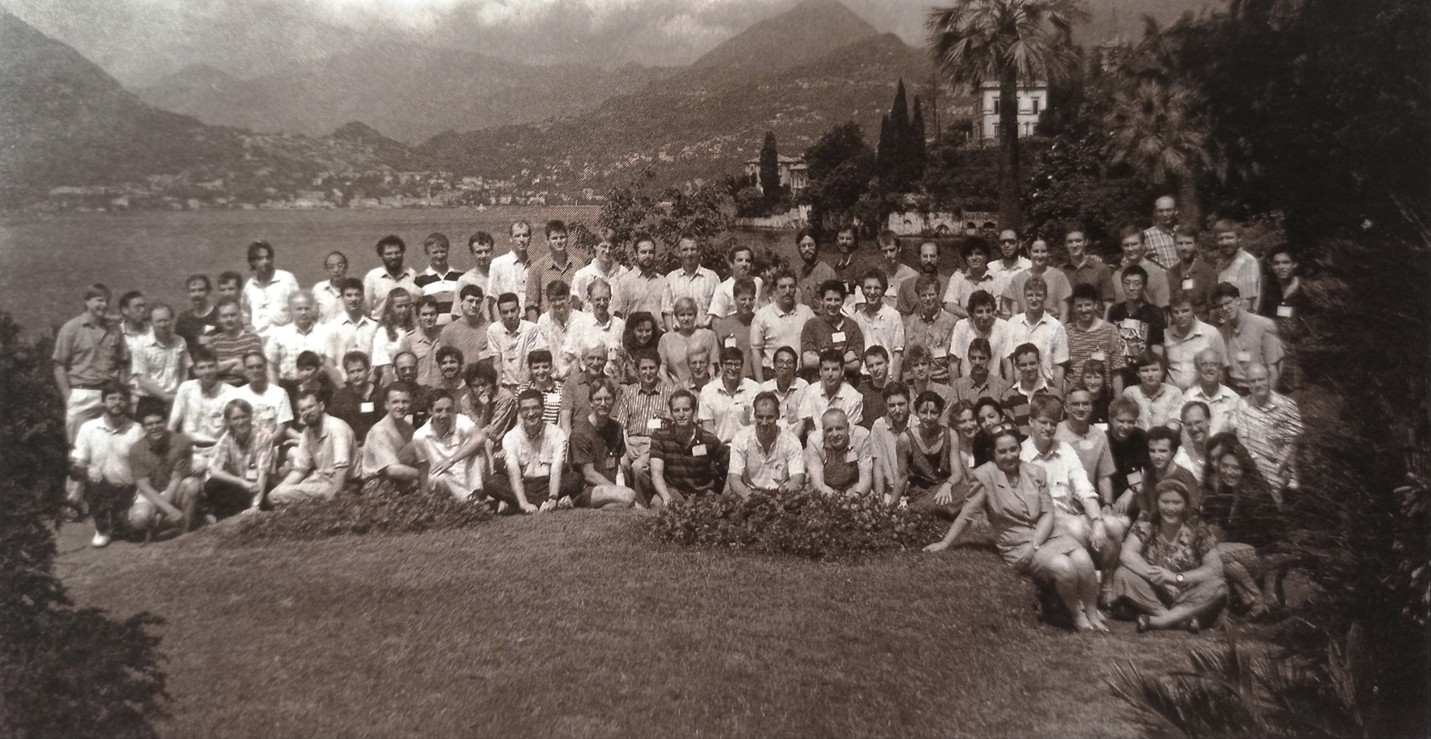}
	\caption{Group photo of the International School of Physics ``Enrico Fermi'' Course CXVIII on ``Laser Manipulation of Atoms and Ions'' that took place in Villa Monastero in Varenna on Lake Como between 9th and 19th July 1991. The course was directed by E. Arimondo, W. D. Phillips and F. Strumia. Copyright Societ\`a Italiana di Fisica (SIF), reproduced with permission.}
	\label{Inguscio:fig:course_CXVIII}
\end{figure}

The Varenna scientific program related to atomic physics eventually shifted from the observation of atomic properties to the active manipulation of matter through laser control, including the production of new kind of matter. Course CXVIII in 1991 (Figure~\ref{Inguscio:fig:course_CXVIII}) served as the definitive turning point toward the laser control of atoms and ions.	This paradigm shift is exemplified by the career of Wolfgang Ketterle, who attended the 1991 school as a student of Herbert Walther on molecular spectroscopy. Profoundly influenced by Alain Aspect’s lectures on sub-recoil Doppler cooling and other methods that later in 1997 brought the Nobel Prize to S. Chu, C. Cohen-Tannoudji and W. D. Phillips~\cite{CohenTannoudji1998}, Ketterle shifted his research focus toward the manipulation of matter, a decision that in foresight facilitated the subsequent creation of new states of matter. 

\begin{figure}[htb]
	\centering
	\includegraphics[width=0.7\textwidth]{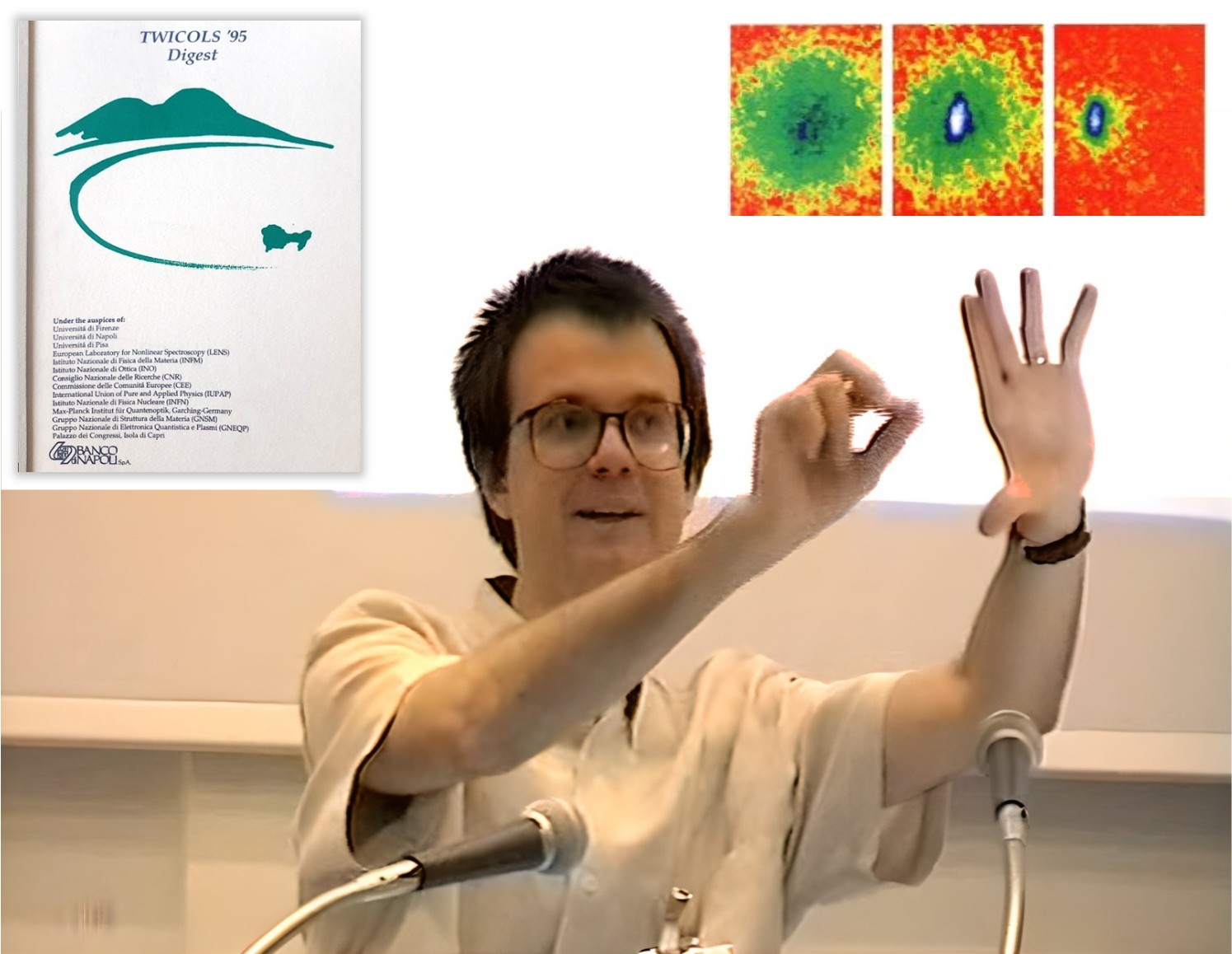}
	\caption{Eric Cornell presenting the achievement of the first Bose-Einstein condensation at the 20th International Conference on Laser Spectroscopy (Capri, 11th - 16th July 1995).}
	\label{Inguscio:fig:BEC_TWICOLS}
\end{figure}

The advancement in cooling techniques eventually led to reaching the threshold of Bose-Einstein condensation (BEC)~\cite{Bose1924, Einstein1924, Anderson1995, Davis1995}, a milestone that inaugurated the era of ultra-cold matter research, where the de Broglie wavelength reaches macroscopic proportions. Eric Cornell first presented the experimental achievement of BEC at the 1995 conference in Capri (Figure~\ref{Inguscio:fig:BEC_TWICOLS}). He was 33 years old then, and this shall motivate young researchers to speed up. Several groups raced towards this result, opening the era of ultra-cold matter studies and producing in 2001 the Nobel Prize for C. Wieman, E. Cornell and W. Ketterle~\cite{Cornell2002, Ketterle2002}. 

\begin{figure}[htb]
	\centering
	\includegraphics[width=\textwidth]{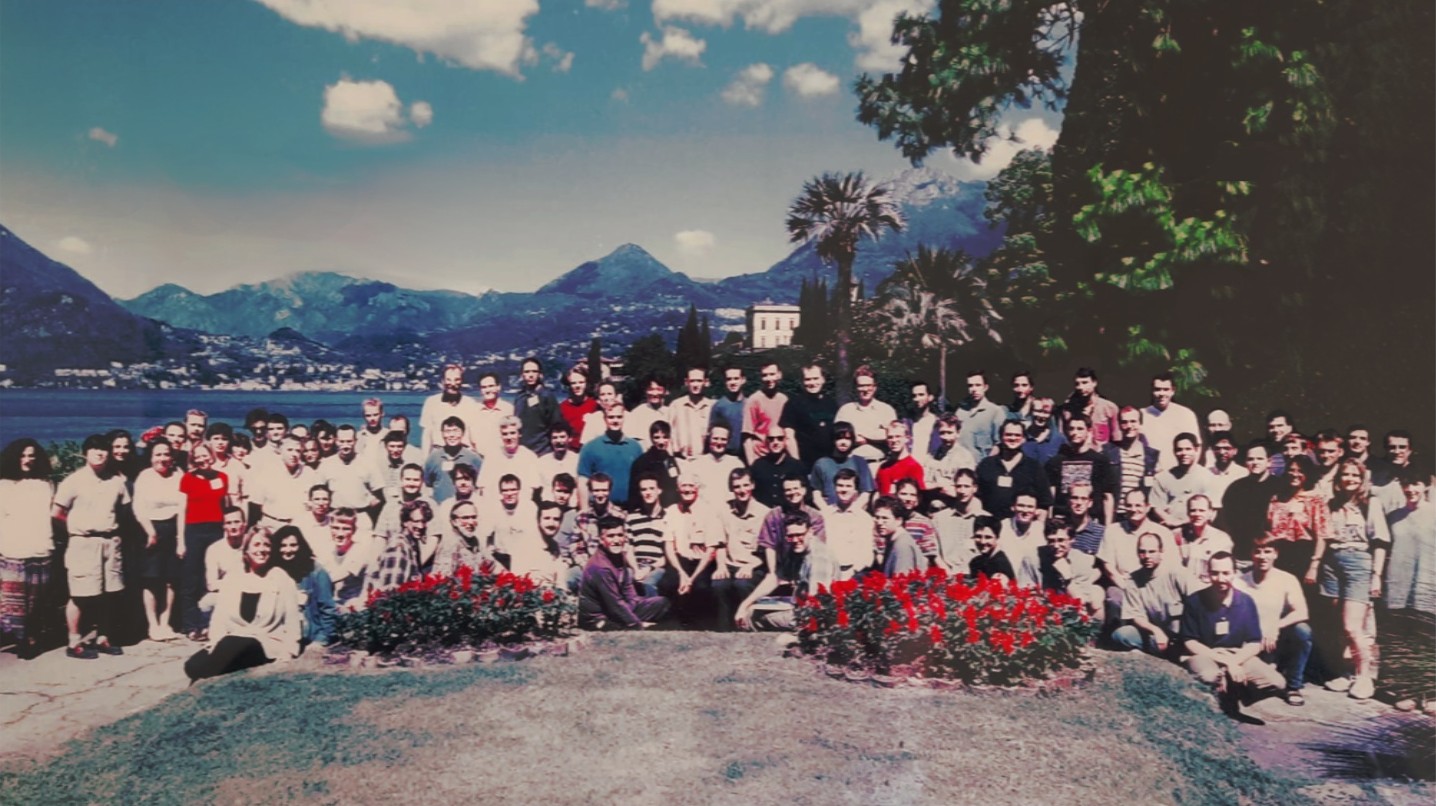}
	\caption{Group photo of the International School of Physics ``Enrico Fermi'' Course CXL on ``Bose-Einstein Condensation in Atomic Gases'' that took place in Villa Monastero in Varenna on Lake Como between 7th and 17th July 1998. The course was directed by M. Inguscio, S. Stringari and C. E. Wieman. Copyright Societ\`a Italiana di Fisica (SIF), reproduced with permission.}
	\label{Inguscio:fig:course_CXL}
\end{figure}

\begin{figure}[htb]
	\centering
	\includegraphics[width=0.8\textwidth]{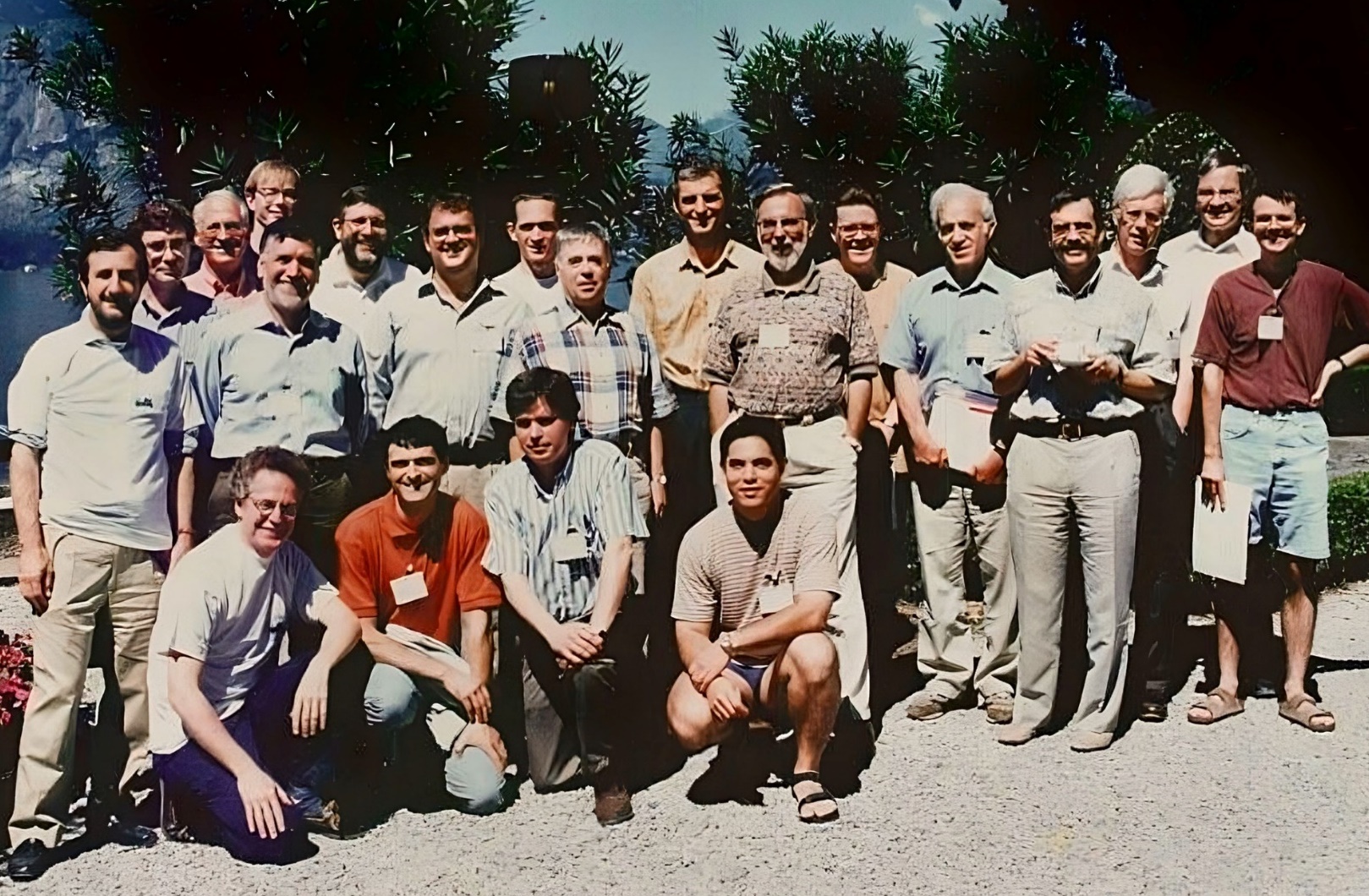}
	\caption{Group photo of several participants to the International School of Physics ``Enrico Fermi'' Course CXL on ``Bose-Einstein Condensation in Atomic Gases'' (1998). Notably, by now six of them received the Nobel Prize: W. D. Phillips (1997), C. Wieman (2001), E. Cornell (2001), W. Ketterle (2001), T. H\"ansch (2005), and A. Aspect (2022). Copyright Societ\`a Italiana di Fisica (SIF), reproduced with permission.}
	\label{Inguscio:fig:course_CXL_nobel}
\end{figure}

Following the initial breakthroughs in the field, Varenna's course CXL addressed in 1998 the BEC physics in atomic gases (Figure~\ref{Inguscio:fig:course_CXL}). This focus on emerging frontiers continued through subsequent schools dedicated to degenerate Fermi gases (course CXL), quantum matter (course 191), and the complexities of quantum mixtures (course 211). This school was attended by a young Immanuel Bloch, whose journey from student to distinguished lecturer illustrates a common growth path supported by Varenna events. Almost everyone has been once a student and we are glad to note that several teachers became Nobel Prize winners (Figure~\ref{Inguscio:fig:course_CXL_nobel}), such as W. D. Phillips (1997), E. Cornell (2001), C. Wieman (2001), W. Ketterle (2001), T. H\"ansch (2005), and A. Aspect (2022), underscoring the school's central role in the global scientific community. The theoretical foundations of course CXL were significantly enriched by the participation of Lev Pitaevskii, a master of ultralow-temperature physics and a key architect of the Gross–Pitaevskii equation, who was very warmly hosted in Italy in Trento.

\begin{figure}[htb]
	\centering
	\includegraphics[width=0.7\textwidth]{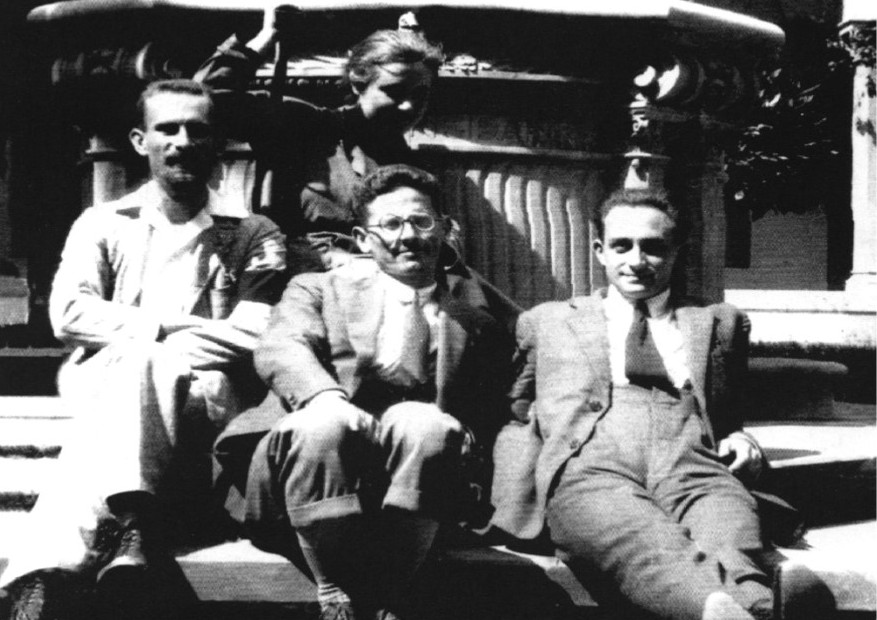}
	\caption{Photo of (from left to right) Franco Rasetti, Rita Brunetti (behind), Nello Carrara, and Enrico Fermi sitting in 1926 in front of the well at the Institute of Physics in Arcetri, Florence. Courtesy of Sapienza Universit\`a di Roma, Archivio del Dipartimento di Fisica, Fondo Edoardo Amaldi.}
	\label{Inguscio:fig:Fermi_Arcetri}
\end{figure}

The exploration of quantum matter statistics necessarily leads to the study of fermions. While Fermi’s final contributions were delivered in Varenna in 1954, his foundational work began decades earlier in Florence at Arcetri, a site of great historical significance for physics, starting from Galileo Galilei who spent close-by the last part of his life. Here at the Institute of Physics, at less than twenty-five years of age (Figure~\ref{Inguscio:fig:Fermi_Arcetri}), Fermi authored his seminal 1926 paper on the statistics governing non-interacting particles subject to the Pauli exclusion principle~\cite{Fermi1926}. This work is arguably his most important theoretical legacy, providing the conceptual bedrock for understanding many physical phenomena related to fermions across the universe.

\begin{figure}[htb]
	\centering
	\includegraphics[width=\textwidth]{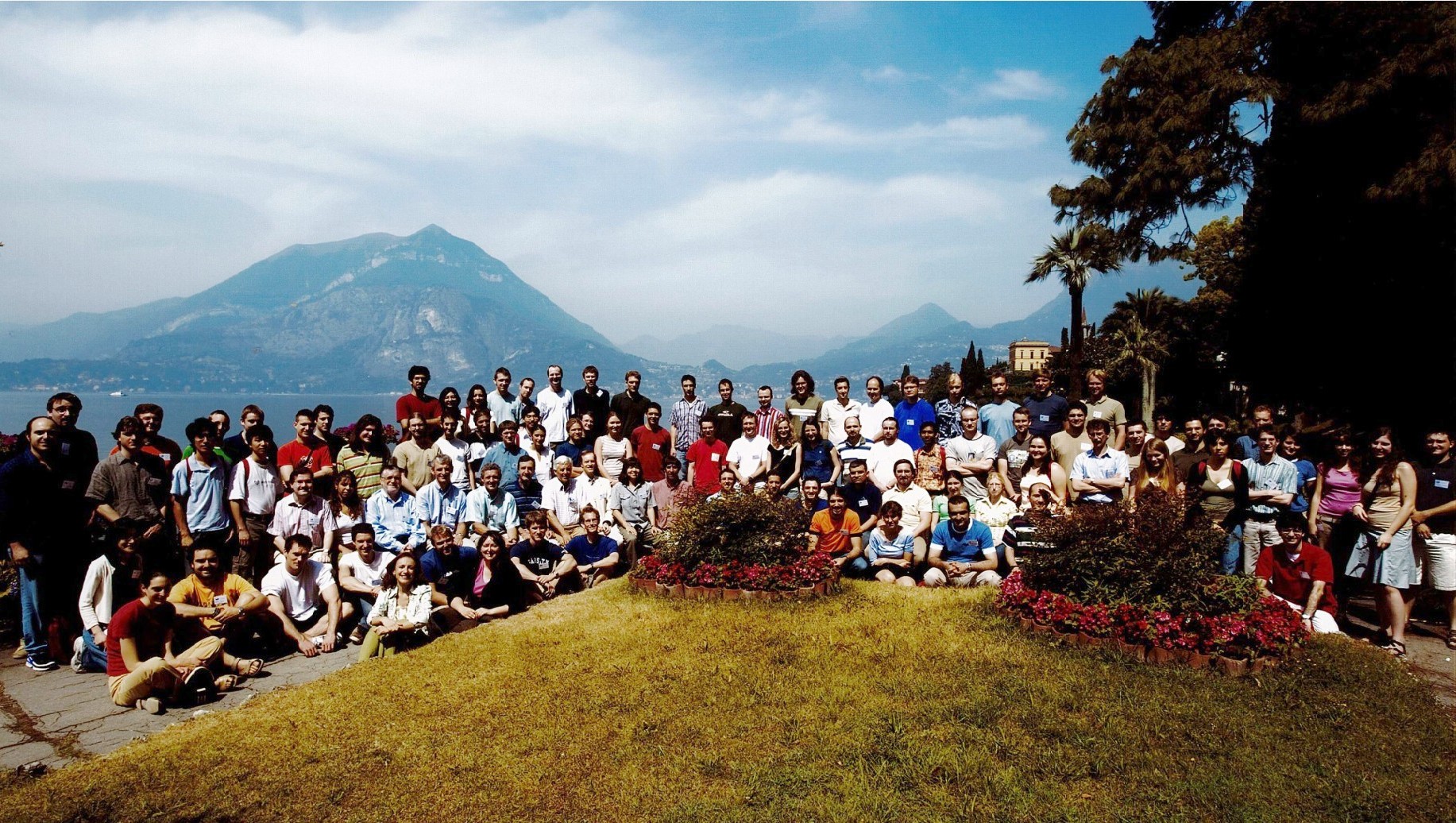}
	\caption{Group photo of the International School of Physics ``Enrico Fermi'' Course CLXIV on ``Ultra-cold Fermi Gases'' that took place in Villa Monastero in Varenna on Lake Como between 20th and 30th June 2006. The course was directed by M. Inguscio, W. Ketterle and C. Salomon. Copyright Societ\`a Italiana di Fisica (SIF), reproduced with permission.}
	\label{Inguscio:fig:course_CLXIV}
\end{figure}

\begin{figure}[htb]
	\centering
	\includegraphics[width=0.7\textwidth]{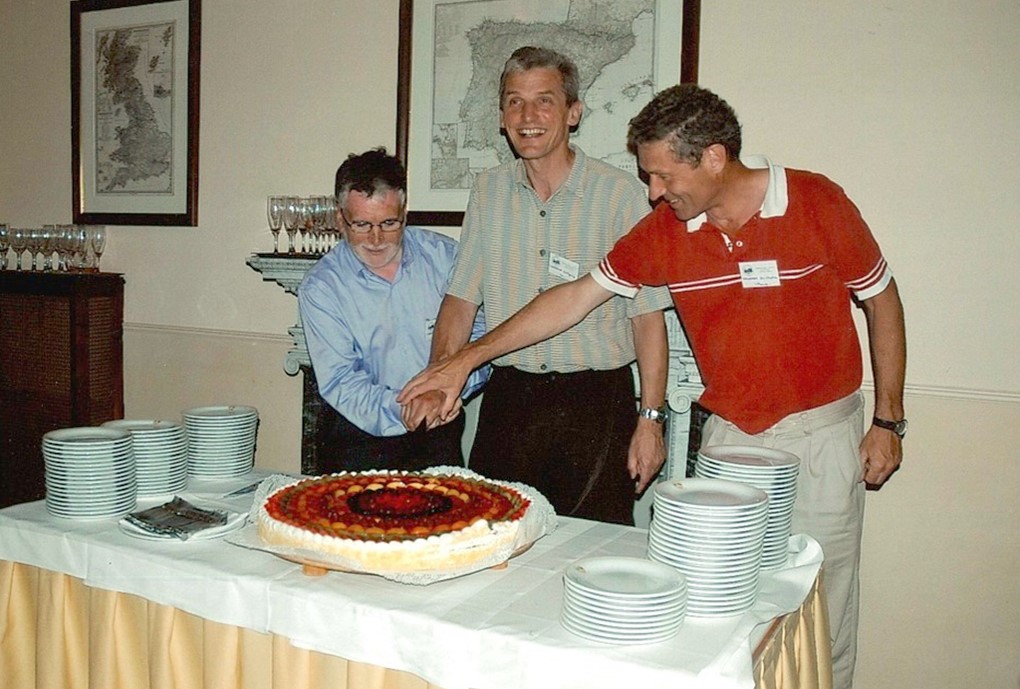}
	\caption{Photo of (from left to right) M. Inguscio, W. Ketterle and C. Salomon at the International School of Physics ``Enrico Fermi'' Course CLXIV on ``Ultra-cold Fermi Gases'' cutting the commemorative cake for the 80th anniversary of Fermi’s 1926 statistics.}
	\label{Inguscio:fig:course_CLXIV_celebration}
\end{figure}

A set of key fundamental questions arose on the behavior of fermionic and bosonic species at ultralow temperatures, leading to significant theoretical and experimental achievements. Notably, in the immediate vicinity of the historic Arcetri site, after the achievement of Bose-Einstein condensation of~\isotope[87]{Rb}~\cite{Fort2000} and the first condensation of~\isotope[41]{K}~\cite{Modugno2001}, one of the earliest degenerate Fermi gases was produced~\cite{Roati2002}. Course CLXIV was convened in 2006 (Figure~\ref{Inguscio:fig:course_CLXIV}) to focus on ultracold Fermi gases and also served as a commemorative event for the 80th anniversary of Fermi’s 1926 statistics (Figure~\ref{Inguscio:fig:course_CLXIV_celebration}). Interestingly, Immanuel Bloch returned here as a lecturer, discussing the first experimental demonstration of the Mott insulator-to-superfluid transition~\cite{Greiner2002} and follow-up studies, which represent a fundamental result that confirmed the viability of using ultracold atoms as effective simulators for complex many-body systems. It is historically significant that Nevill F. Mott, who first described this transition~\cite{Mott1968}, was a participant at the Varenna School in 1957, exactly two decades before being awarded the Nobel Prize in 1977 together with Philipp W. Anderson and John H. Van Vleck.

\begin{figure}[htb]
	\centering
	\includegraphics[width=\textwidth]{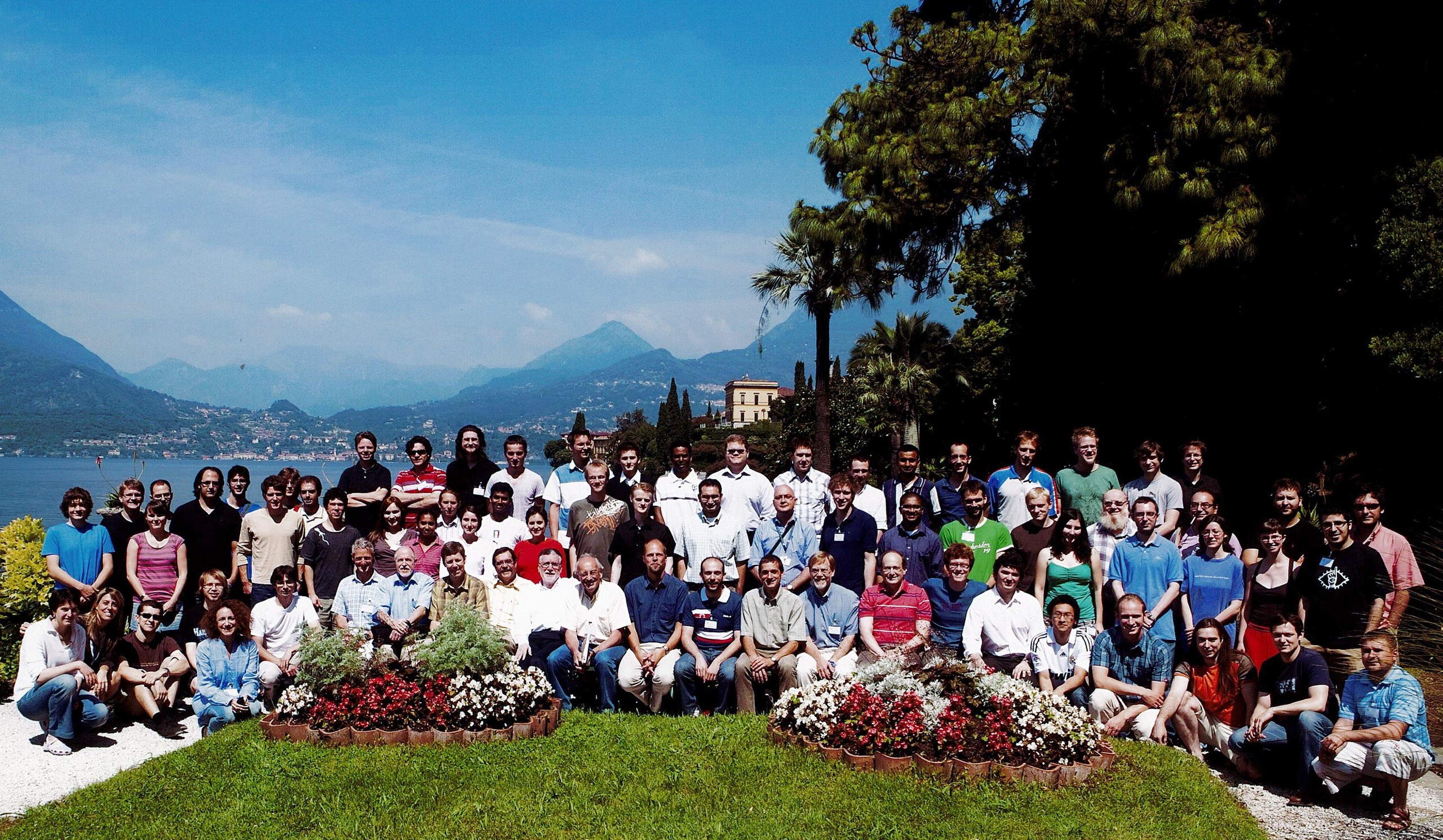}
	\caption{Group photo of the International School of Physics ``Enrico Fermi'' Course CLXXIII on ``Nano Optics and Atomics: Transport of Light and Matter Waves'' that took place in Villa Monastero in Varenna on Lake Como between 23rd June and 3rd July 2009. The course was directed by R. Kaiser and D. S. Wiersma, with the support of the scientific secretary L. Fallani. Copyright Societ\`a Italiana di Fisica (SIF), reproduced with permission.}
	\label{Inguscio:fig:course_CLXXIII}
\end{figure}

\begin{figure}[htb]
	\centering
	\includegraphics[width=\textwidth]{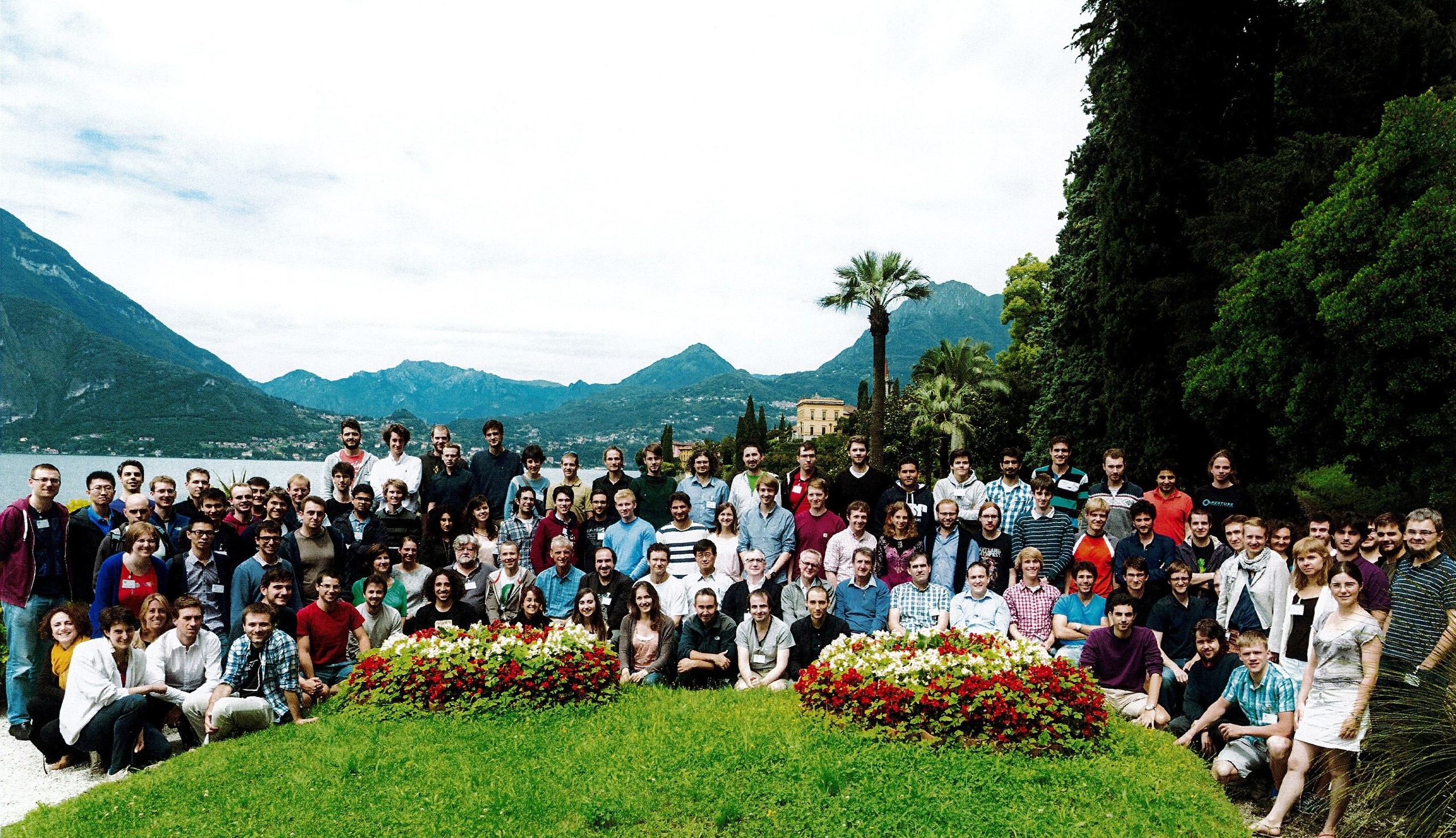}
	\caption{Group photo of the International School of Physics ``Enrico Fermi'' Course 191 on ``Quantum Matter at Ultralow Temperatures'' that took place in Villa Monastero in Varenna on Lake Como between 7th and 15th July 2014. The course was directed by M. Inguscio, W. Ketterle and S. Stringari, with the support of the scientific secretary G. Roati. Copyright Societ\`a Italiana di Fisica (SIF), reproduced with permission.}
	\label{Inguscio:fig:course_191}
\end{figure}

\begin{figure}[htb]
	\centering
	\includegraphics[width=\textwidth]{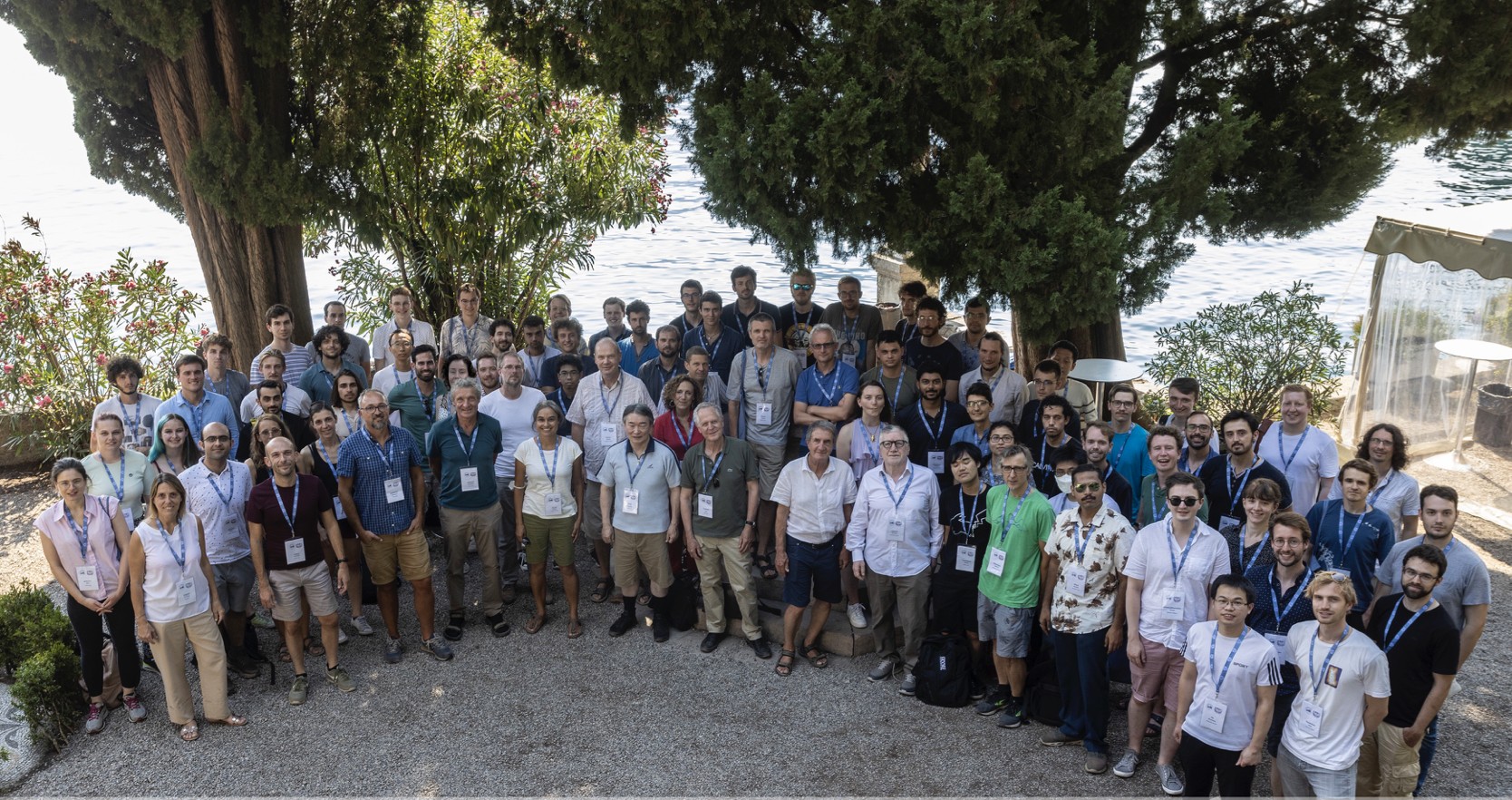}
	\caption{Group photo of the International School of Physics ``Enrico Fermi'' Course 211 on ``Quantum Mixtures With Ultra-Cold Atoms'' that took place in Villa Monastero in Varenna on Lake Como between 18th and 23rd July 2022. The course was directed by R. Grimm, M. Inguscio and S. Stringari, with the support of the scientific secretary G. Lamporesi. Copyright Societ\`a Italiana di Fisica (SIF), reproduced with permission.}
	\label{Inguscio:fig:course_211}
\end{figure}

Next, the research scope expanded to include the study of transport phenomena in disordered systems. In course CLXXIII (Figure~\ref{Inguscio:fig:course_CLXXIII}), Leonardo Fallani addressed the simulation of Anderson localization~\cite{Anderson1958} using atoms in disordered optical potentials. Alain Aspect described this period as a ``friendly competition'' between the French and Italian scientific communities that accelerated progress in the field. Later in 2014, the dual role in organizational support and scientific innovation of Giacomo Roati was instrumental to the success of course 191 (Figure~\ref{Inguscio:fig:course_191}) that covered the major new directions in ultracold quantum matter research, including Bose-Fermi mixtures, the impact of strong and long range interactions, quantum gas microscopy and simulation of quantum magnetism in optical lattices, disorder, reduced dimensionality and non-trivial topological properties. The most recent developments, as explored in course 211 (Figure~\ref{Inguscio:fig:course_211}), have centered on quantum mixtures of different atom species, expanding from alkali metals to many more types, in particular alkaline earths and lanthanides. Their richer electronic structure offers unique advantages for high-precision metrology, realizing state-of-the-art optical clocks, and for simulating advanced quantum models, like SU($N$) quantum magnetism.

\begin{figure}[htb]
	\centering
	\includegraphics[width=\textwidth]{Figures_ch_Inguscio/Fermi_Computer.jpg}
	\caption{Fermi and computing. \textbf{(a)} Enrico Fermi's letter to Enrico Avanzi, Rector of Pisa University (11th August 1954). \textbf{(b)} Photo of the first Italian scientific calculator C.E.P. (Calcolatrice Elettronica Pisana) built in 1961. Courtesy of the University of Pisa.}
	\label{Inguscio:fig:Fermi_Computer}
\end{figure}

The significant efforts invested over the years into finely controlling ultracold matter to realize analog quantum simulations of selected hamiltonians are contributing also towards the goal of bringing about digital quantum computation. A broad range of physical platforms are being explored by many research teams across the world and this diffuse and bottom-up approach resonates with Fermi's vision of scientific inquiry. Fermi became very familiar with the nascent field of electronic computing in the course of his work in the United States of America, learning in 1952 to work with the Maniac computer. During the 1954 school, he was asked to advise on the allocation of a significant research budget for a new endeavor in Italy. Rather than recommending a second particle accelerator beyond the one that was planned to be built in Frascati, Fermi advocated for the indigenous construction of a computer at the University of Pisa, in particular by addressing the letter displayed in Figure~\ref{Inguscio:fig:Fermi_Computer}a to Enrico Avanzi, Rector of Pisa University. He argued that the act of building the machine was scientifically superior to simply purchasing one, as it ensured a fundamental mastery of the technology and a true advancement for a research center, a prerequisite for conducting cutting edge scientific inquiry across multiple disciplines - a philosophy that mirrors contemporary efforts in constructing quantum computers. Embracing Fermi’s advice, the construction of the C.E.P. (Calcolatrice Elettronica Pisana) commenced, eventually resulting in 1961 in the machine shown in Figure~\ref{Inguscio:fig:Fermi_Computer}b~\cite{Cignoni2020, Cignoni2025}. The project benefited from significant institutional support that recognized the importance of science, including tax exemptions for the required specialized vacuum tubes since their application was purely scientific rather than commercial. 

\begin{figure}[htb]
	\centering
	\includegraphics[width=\textwidth]{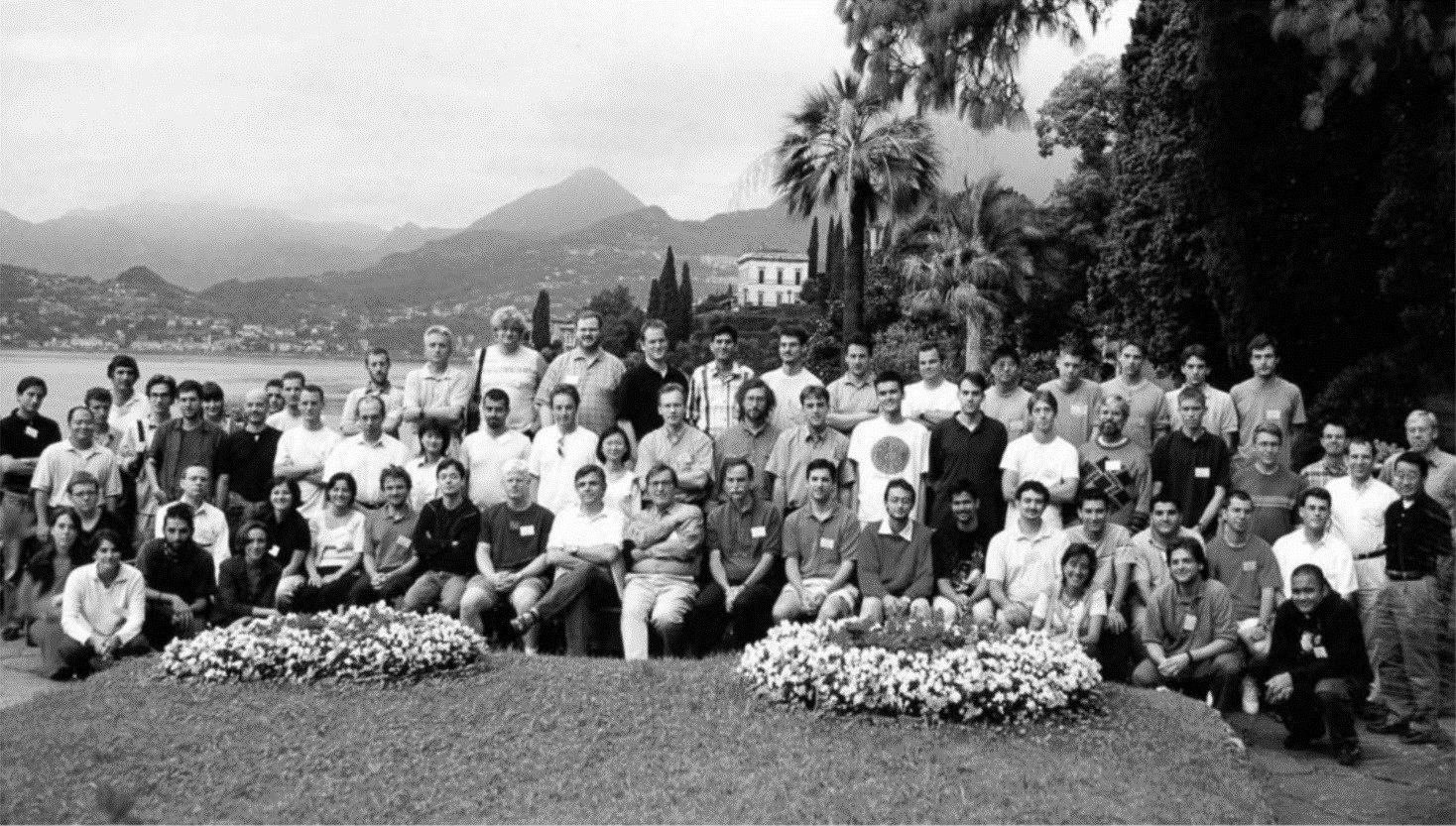}
	\caption{Group photo of the International School of Physics ``Enrico Fermi'' Course CXLVIII on ``Experimental Quantum Computation and Information'' that took place in Villa Monastero in Varenna on Lake Como between 17th and 27th July 2001. The course was directed by F. De Martini and C. Monroe. Copyright Societ\`a Italiana di Fisica (SIF), reproduced with permission.}
	\label{Inguscio:fig:course_CXLVIII}
\end{figure}

\begin{figure}[htb]
	\centering
	\includegraphics[width=\textwidth]{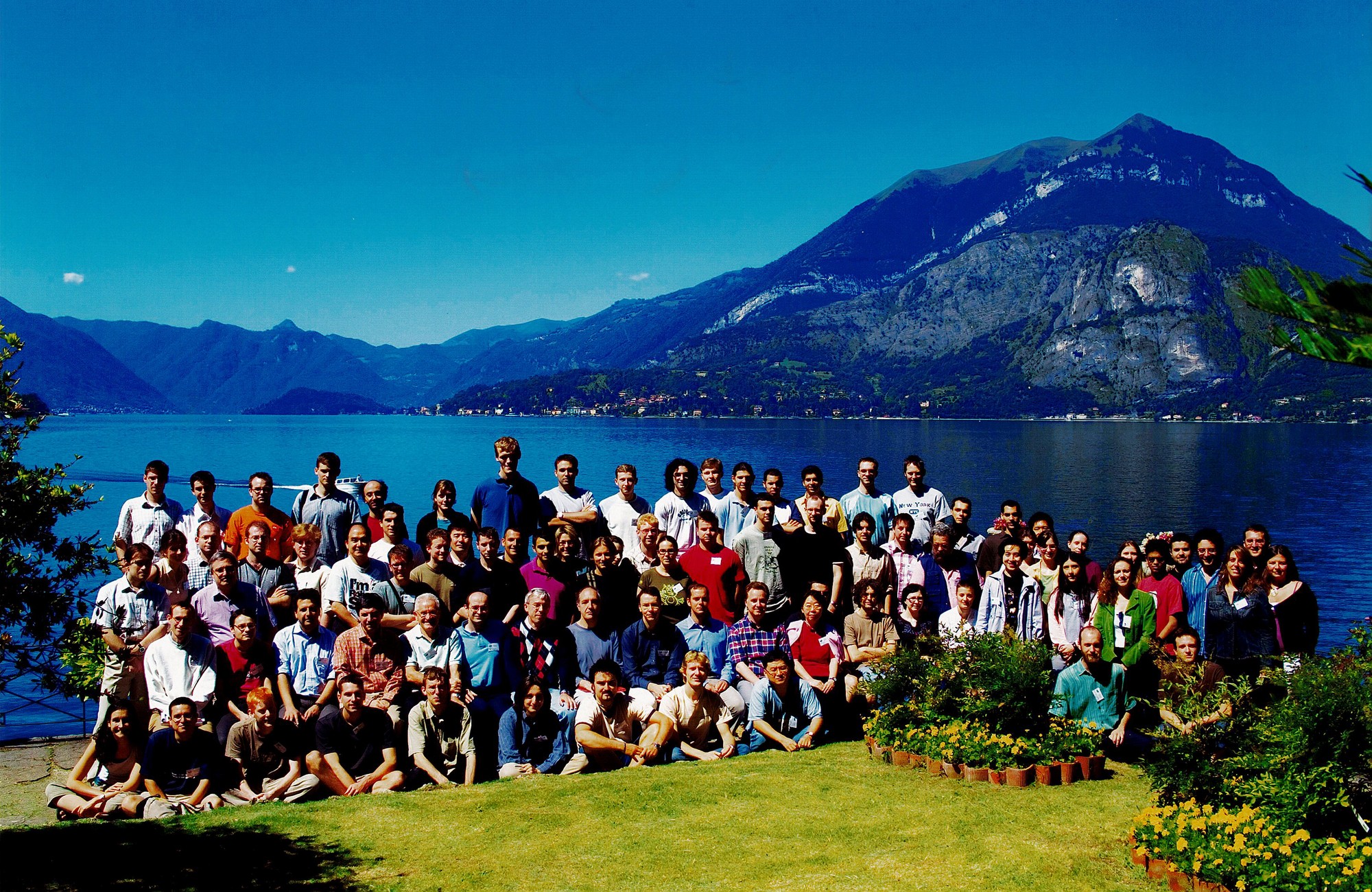}
	\caption{Group photo of the International School of Physics ``Enrico Fermi'' Course 162 on ``Quantum Computers, Algorithms and Chaos'' that took place in Villa Monastero in Varenna on Lake Como between 5th and 15th July 2005. The course was directed by G. Casati, D. L. Shepelyansky and P. Zoller. Copyright Societ\`a Italiana di Fisica (SIF), reproduced with permission.}
	\label{Inguscio:fig:course_162}
\end{figure}

The following technological trajectory - from vacuum-tubes, pentodes, diodes, through progressively miniaturized transistors and chips, onto the manipulation of individual atoms to create a quantum computer - reflects a fundamental shift in information science, captured by a far viewing observation of one director of the 1991 Varenna course: ``Quantum information is a radical departure in information.''. The modern capability to manipulate atoms with optical tweezer arrays is largely due to the work of Antoine Browaeys~\cite{Beugnon2007, Barredo2016, Labuhn2016, Bernien2017}, who is closely linked to the Varenna tradition, having attended as a student the course CXLVIII on ``Experimental Quantum Computation and Information'' in 2001 (Figure~\ref{Inguscio:fig:course_CXLVIII}). This topic was further explored in the course 162 on ``Quantum Computers, Algorithms and Chaos'' carried out between 5th and 15th July 2005 and directed by G. Casati, D. L. Shepelyansky and P. Zoller (Figure~\ref{Inguscio:fig:course_162}), with the support of the scientific secretary G. Benenti.

\begin{figure}[htb]
	\centering
	\includegraphics[width=0.7\textwidth]{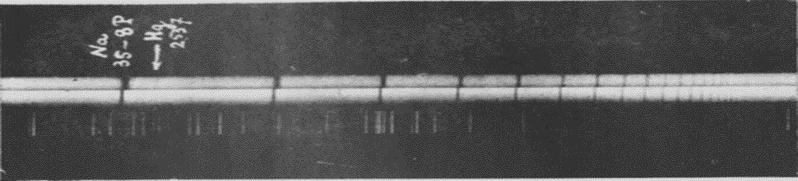}
	\caption{Photo of two side-by-side~\isotope{Na} absorption spectrum lines in presence of two concentrations of~\isotope{H}$_2$ buffer gas, revealing a shift that does not affect the \isotope{Hg} line. Figure reproduced with permission from~\cite{Amaldi1934}.}
	\label{Inguscio:fig:Fermi_Rydberg}
\end{figure}

Beyond single atom trapping and manipulation in optical tweezers, a second capability is crucial towards realizing effective quantum computers on a neutral atom-based platform, namely attaining and controlling long-range interactions between particles hosted in different traps at micrometer-range separations, creating entanglement between them. This is made possible by exciting the atoms from the ground to a Rydberg state, enabling the implementation of several types of quantum gates~\cite{Graham2022, Bluvstein2023}. Interestingly, Fermi's legacy is interwoven with these highly excited atoms, to which he colloquially referred as \textit{atomi ciccioni} (ultra-fat atoms), as recounted by his student Edoardo Amaldi. In 1934, Amaldi shared with Fermi that they observed in the lab an unexpected shift of the transition frequencies of~\isotope{Na} atoms under high~\isotope{H}$_2$ buffer gas pressure~\cite{Amaldi1934} (Figure~\ref{Inguscio:fig:Fermi_Rydberg}). Fermi explained that this effect was due to the alkali atoms being highly excited~\cite{Fermi1934}, which greatly affects their physical properties, including a significant expansion of their electronic orbits. Amaldi recalls that Fermi employed a visual metaphor to explain what a Rydberg atom was: ``Oh, these are atoms that are so large, their orbits are so large, they’re like fat atoms (\textit{ciccioni}).'' Furthermore, Fermi introduced in that paper the concept of the \textit{scattering length} which is now an important and common tool in the field of ultracold atoms. This clear and visual approach to complex physics continues to inspire contemporary research in quantum simulation and computation.

\begin{figure}[htb]
	\centering
	\includegraphics[width=\textwidth]{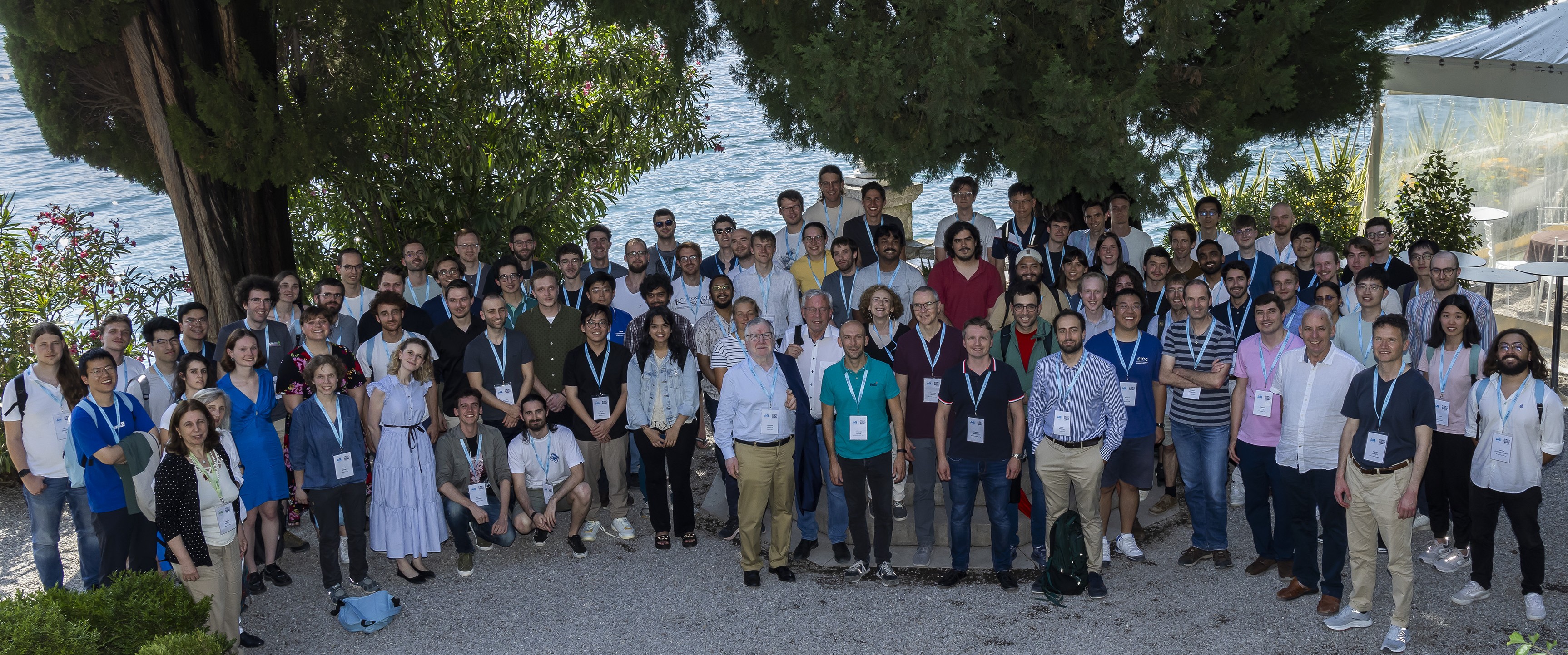}
	\caption{Group photo of the International School of Physics ``Enrico Fermi'' Course 214 on ``Quantum Computers and Simulators with Atoms'' that took place in Villa Monastero in Varenna on Lake Como between 5th and 14th July 2024. The course was directed by I. Bloch, A. Browaeys and L. Fallani, with the support of the scientific secretary V. Gavryusev. Copyright Societ\`a Italiana di Fisica (SIF), reproduced with permission.}
	\label{Inguscio:fig:course_214}
\end{figure}

All of this has happened here in Villa Monastero in Varenna, and we believe it is not by chance. There is something very special about this place. Not only the environment, the lake and the beauty are stimulating, but also the history, the continuity, the culture of curiosity and respect for science and between people that carry it out. The participants in the course 214 on ``Quantum Computers and Simulators with Atoms'' held in 2024 (Figure~\ref{Inguscio:fig:course_214}) represent the latest link in a historical chain of explorers that extends back to Fermi. The responsibility for future discovery and the continued vibrancy of the scientific endeavor, while preserving its connection to humanity and nurturing human relationships, now resides with this new generation of researchers.

\begin{acknowledgement}
V. Gavryusev acknowledges support, in the context of the National Recovery and Resilience Plan and Next Generation EU, from M4C2 investment 1.2 project MicroSpinEnergy.
\end{acknowledgement}

\ethics{Competing Interests}{The authors have no conflicts of interest to declare that are relevant to the content of this chapter.}

\eject

%
%
\bibliographystyle{spphys.bst}
\bibliography{references_ch_Inguscio}

\end{document}